\documentclass[aps,pre,twocolumn,amsfonts,amssymb,amsmath,floatfix,showpacs]{revtex4}
\usepackage[dvips]{graphicx}
\usepackage{color}

\newcommand{\Heav}{\Theta}             
\newcommand{\df}[2]{\frac{\partial #1}{\partial #2}} 
\newcommand{\duration}{\Delta}         
\newcommand{\dt}{\Delta t}             
\newcommand{\dx}{\Delta x}             
\newcommand{\omegaf}{\omega_f}         
\newcommand{\omegas}{\omega_0}         
\newcommand{\omegat}{\bar{\omega}_0}   
\newcommand{\omegac}{\tilde{\omega}_0} 
\newcommand{\tdelay}{t_{\mathrm{delay}}}
\newcommand{\tr}{t^r}                   
\newcommand{\ts}{t^s}                   
\newcommand{\uc}{u_*}                   
\newcommand{\untuning}{\delta}          
\newcommand{\xr}{x_r}                   
\newcommand{\yr}{y_r}                   
\newcommand{\zr}{z_r}                   

\newcommand{\eg}{\textit{e.g.}}
\newcommand{\etal}{\textit{et al.}}
\newcommand{\Fig}[1]{Fig.~\ref{#1}}
\newcommand{\fig}[1]{fig.~\ref{#1}}
\newcommand{\ie}{\textit{i.e.}}

\newcommand{\dblfigure}[3]{\begin{figure*}#1\caption[]{#2}\label{#3}\end{figure*}}
\newcommand{\sglfigure}[3]{\begin{figure}#1\caption[]{#2}\label{#3}\end{figure}}

\begin{document}

\title{Control of scroll wave turbulence using resonant perturbations}
\author{S. W. Morgan}
\author{I. V. Biktasheva}
\author{V. N. Biktashev}
\affiliation{Department of Mathematical Sciences, University of Liverpool, Liverpool L69 7ZL, UK}
\affiliation{Department of Computer Science, University of Liverpool, Liverpool L69 3BX, UK}

\date{\today}

\begin{abstract}
  Turbulence of scroll waves is a sort of spatio-temporal chaos
  that exists in three-dimensional excitable media. Cardiac tissue and
  the Belousov-Zhabotinsky reaction are examples of such media. In
  cardiac tissue, chaotic behaviour is believed to underlie fibrillation
  which, without intervention, precedes cardiac death. In this study we
  investigate suppression of the turbulence using stimulation of
  two different types, ``modulation of excitability'' and ``extra
  transmembrane current''.  With cardiac defibrillation in mind, we
  used a single pulse as well as repetitive extra current with
  both constant and feedback controlled frequency. We show that
  turbulence can be terminated using either a resonant modulation
  of excitability or a resonant extra current. The turbulence is
  terminated with much higher probability using a resonant frequency
  perturbation than a non-resonant one. Suppression of the turbulence
  using a resonant frequency is up to fifty times faster than
  using a non-resonant frequency, in both the modulation of
  excitability and the extra current modes. We also demonstrate
  that resonant perturbation requires strength one order of
  magnitude lower than that of a single pulse, which is currently used
  in clinical practice to terminate cardiac fibrillation. Our results
  provide a robust method of controlling complex chaotic spatio-temporal
  processes. Resonant drift of spiral waves has been studied extensively
  in two dimensions, however, these results show for the first
  time that it also works in three dimensions, despite the complex
  nature of the scroll wave turbulence.
\end{abstract}

\maketitle

\section{Introduction}

Turbulence of scroll waves is a sort of spatiotemporal chaos
that is observed in some three-dimensional (3D) excitable
media~\cite{%
  Brazhnik-etal-1987,%
  Biktashev-etal-1994,%
  Winfree-1994,%
  Biktashev-1998,
  Fenton-Karma-1998-PRL,%
  Fenton-Karma-1998-Chaos,%
  Alonso-etal-2003,
  Alonso-etal-2006-JPC,%
  Alonso-Panfilov-2007%
}. In cardiac tissue, such chaotic behaviour is
known as fibrillation \cite{Gray-Jalife-1996} and implies cardiac failure.

The current method for terminating fibrillation in cardiac tissue is
by means of a single electric pulse with a large amplitude. However,
this approach is far from ideal. There are several known side-effects
linked to the administration of the large electric shocks to patients
\cite{Boriani-etal-2005}. Termination of fibrillation using
shocks with a lower amplitude would overcome such problems.

Resonant drift of a spiral wave in two dimensions (2D) has been
observed when one of the parameters of a model of an excitable medium
was changed in time, with the period equal to that of the spiral wave
\cite{Davydov-etal-1988, Agladze-etal-1987}. This phenomenon was later
shown to be generic for reaction-diffusion excitable systems
\cite{Biktashev-Holden-1993,Biktashev-Holden-1995}. Thus it appeared
that resonant drift could be used for moving the spiral wave around
the medium to a boundary where it would terminate.

Numerical experiments with reaction-diffusion models revealed that
when close to boundaries, the period of the spiral wave changes, thus
destroying the resonance in such a way that the drift trajectory turns
away from the boundary. The main reason for this unruly behaviour is
``resonant repulsion'', caused by
the untuning of the resonance between the spiral and the
perturbation, due to the variation of the spiral's own frequency
\cite{Biktashev-Holden-1993,Biktashev-Holden-1995}. This can be rectified by adjusting the
frequency of the external forcing accordingly, based on some kind of
feedback obtained from the re-entry itself. Feedback control of the
resonant drift has been shown to overcome repulsion from the
boundaries and inhomogeneities \cite{Biktashev-Holden-1994,Biktashev-Holden-1995}.
The same method can also eliminate
multiple spiral waves, thus demonstrating that multiplicity of 
re-entrant sources in fibrillation is not in itself an obstacle
for low-voltage defibrillation by this method~\cite{Biktashev-Holden-1995}.

Chambers of a heart, particularly ventricles, are 3D, so termination
of scroll waves should be studied. Scroll wave turbulence presents a
new challenge for resonant drift control, since here we are dealing
not only with multiple sources, but sources that tend to multiply.

A scroll wave rotates around a central filament.  Depending on the
parameters of the medium, a filament of a circular shape may gradually
contract or expand with time
\cite{Panfilov-Rudenko-1987,Brazhnik-etal-1987}.  For filaments of
arbitrary shape, this property translates into ``filament
tension''~\cite{Biktashev-etal-1994}.

In excitable media, if a circular filament contracts, then a filament
of any shape in the same medium has ``positive tension'' and will
shorten with time. Scroll waves with positive filament tension
therefore either collapse or stabilise to a straight shape.  In a
bounded medium this can lead to the self-termination of the scroll.

If a circular filament expands, then any filament shape is unstable as
the filament has ``negative tension'' and will tend to lengthen.  It
was therefore
conjectured~\cite{Brazhnik-etal-1987,Biktashev-etal-1994,Winfree-1994}
and subsequently demonstrated~\cite{%
  Biktashev-1998,%
  Alonso-etal-2003,%
  Alonso-etal-2006-JPC,%
  Alonso-Panfilov-2007%
} that
such excitable media of sufficiently big size should support
``turbulence'' of scroll waves, where the scroll filaments grow,
spontaneously bend, and break up to fragments upon collision with
boundaries and with each other. Negative filament tension is not the
only mechanism of scroll wave turbulence: similar behaviour may occur
due to non-uniform anisotropy of diffusivity, like that found in a
ventricular wall~\cite{Fenton-Karma-1998-PRL,Fenton-Karma-1998-Chaos}.

Alonso \etal\ \cite{Alonso-etal-2003} considered the effect of
applying a periodic \emph{non-resonant} forcing on scroll wave
turbulence produced by negative filament tension.  By numerical
simulations of the Barkley model, they showed that periodic modulation
of the medium's excitability with constant frequency higher than the
frequency of the scroll waves can control the turbulence in the
medium. They went on to propose a theory of this effect, based on the
``kinematic description'' of the scroll
waves~\cite{Alonso-etal-2006-Chaos}. Their interpretation is that
faster-than-resonant stimulation can effectively change the filament
tension from negative to positive, thus disrupting the mechanism
supporting the multiplication of scroll flaments, and compelling them
to collapse.

However, to our knowledge, the possibility of eliminating scroll
turbulence by \emph{resonant} stimulation has not been been
investigated so far.  We set this task for the present study.

In this study we compare suppression of the scroll wave turbulence
using (1) modulation of the medium's excitability, as used by Alonso
\etal\ in \cite{Alonso-etal-2003}, to enable comparison of our
resonant forcing results with their non-resonant forcing, and (2) an
``extra transmembrane current'' forcing, as in
\cite{Biktashev-Holden-1995}. Keeping in mind single shock cardiac
defibrillation used in clinical practice, we also compare repetitive
external forcing of constant and feedback controlled frequencies to a
single pulse extra current forcing.

Our results show that resonant perturbation ensures the quickest
termination of the scroll wave turbulence. Feedback controlled
external perturbation was as effective as constant frequency resonant
perturbation, but offers the advantage of not having to know the
correct frequency \textit{a priori}. The resonant and
feedback-controlled forcing suppress the 3D turbulence using
amplitudes one order of magnitude lower than that of a single pulse
currently used in clinical practice to terminate fibrillation.

\section{Methodology}

\subsection{Governing Equations}

The 3D numerical simulations presented here were performed using the
Barkley model of excitable media \cite{Barkley-1991},
\begin{eqnarray}
\df{u}{t} & = & \frac{1}{\epsilon} u(1-u)\left(u - \frac{v+b(t)}{a} \right) + \nabla^{2} u + h(t),\\
\df{v}{t} & = & u-v,
\end{eqnarray}
where $\epsilon$ is a small parameter $\epsilon \ll 1$ characterising
mutual time scales of the fast $u$ and slow $v$ variables, and $a$ and
$b$ specify the kinetic properties of the system. Parameter $b$
determines the excitation threshold and thus controls the excitability
of the medium.  The term $h(t)$ represents an ``extra transmembrane
current''.

\subsection{Numerical Methods}

For numerical simulations, we used EZscroll software by Barkley \etal\
\cite{Barkley-1991,Dowle-etal-1997}, modified appropriately to
describe the stimulation. We used a 19-points finite difference
approximation of the Laplacian with equal discretization steps $\dx$
in all three spatial directions, and an implicit first order Euler
time stepping with a time step $\dt$.  Simulations were run in a box
$(x,y,z)\in[0,L]^3$, with Neumann boundary conditions.  In most
simulations, we used $\dx=2/3$, $\dt=1/30$ and $L=60$.

In all simulations the model parameters were chosen as $a=1.1$,
$b=0.19$ (or the average value of $b(t)$ when it varied), and
$\epsilon=0.02$, as in \cite{Alonso-etal-2003}.  At this set of
parameters a scroll wave will have negative filament tension.

The choice of parameters was the same as in \cite{Alonso-etal-2003} to
allow comparison, with the exception of the discretization steps. We
used cruder discretization steps, which allowed us to perform more
simulations within reasonable CPU time.  We performed also selected
control simulations with $\dx=0.4$, $\dt=0.01$ and $L=60$, as in
\cite{Alonso-etal-2003}; the results were quantitatively somewhat
different but qualitatively similar (see below for details).

\subsection{Generation of turbulence}

\sglfigure{
  \includegraphics{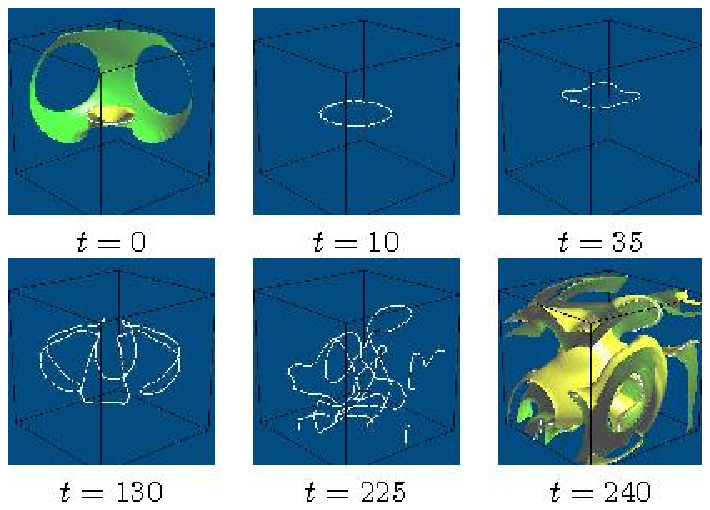}
}{
  (color online)
  Development of scroll wave turbulence from an initial scroll
ring. The white lines show the filaments of the scroll waves.
}{genturb}

The development of scroll wave turbulence is presented in
\fig{genturb}. Starting at $t=0$ with the standard EZScroll scroll
ring initial conditions in an unperturbed medium, the negative tension
of the initial scroll ring caused elongation and bending of the
filament.  Interaction with the boundaries caused the filament to
fragment and soon a complex tangle of many filaments filling the
volume was observed.

The turbulent state of the system was saved at five different times
$t=240, 245, 250, 255, 260$, and then each state used as an initial
condition in our simulations.

\subsection{Resonant Frequency}

\dblfigure{
  \includegraphics{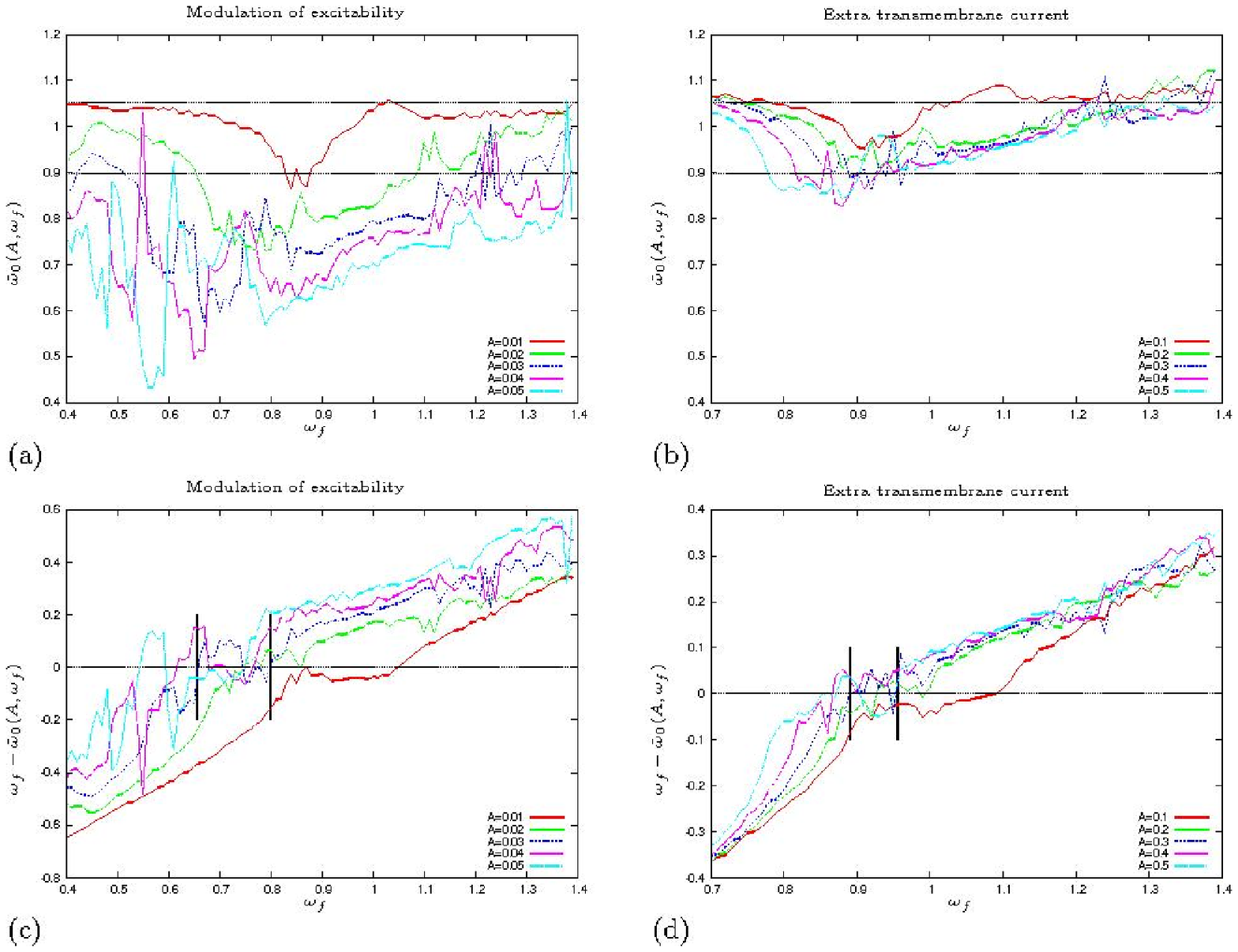}
}{
  (color online) The mean frequency $\omegac(A,\omegaf)$ of the
  perturbed turbulence, measured at the point $(x_{r}, y_{r},
  z_{r})=(0,0,0)$, against forcing frequency $\omegaf$ at different
  forcing amplitudes $A$ due to (a) modulation of the medium's
  excitability and (b) extra transmembrane current forcing.  In (a) and
  (b) the line $\omegat=1.05$ is the mean frequency of the unperturbed
  turbulence, and the line $\omegas=0.9$ is the frequency of the single
  vortex.  The \textbf{resonant windows} can be seen from the deviation
  of the mean frequency $\omegac(A,\omegaf)$ from the forcing frequency
  $\omegaf$ due to (c) modulation of excitability and (d) extra current
  forcing. In (c) and (d) the line $\omegaf-\omegac(A,\omegaf)=0$ is
  drawn to highlight the resonant windows. The vertical brackets
  illustrate the corresponding resonant windows at forcing amplitude
  $A=0.03$ for (c) and $A=0.3$ for (d).
}{figfreq}

There are different ways to define resonance between the forcing and
the turbulence it is aimed to control. We considered three different
frequencies, to which the forcing frequency can be compared:
\begin{itemize}
\item The rotation frequency of an unforced vortex around its filament
$\omegas$. It is also the frequency of a 2D spiral wave in a large
enough medium. It is typically used as the leading-order approximation
in any perturbative theoretical approaches,
as~\cite{Biktashev-etal-1994,Alonso-etal-2003}.
\item The mean frequency of the unforced turbulence $\omegat$. It is
different from $\omegas$ due to interaction of scrolls with each other
and with boundaries. This difference can be significant as this
interaction is the only factor that stops the filaments' growth in
length and number.
\item The mean frequency of the forced turbulence
$\omegac(A,\omegaf)$, which depends on the forcing amplitude $A$, and
the forcing frequency $\omegaf$ (see \fig{figfreq}(a,b)).  By
definition, $\omegac(0,\omegaf)=\omegat$ for any $\omegaf$. The
difference between $\omegac$ and $\omegat$ is less obvious from the
theoretical viewpoint than the difference between $\omegat$ and
$\omegas$, as the theory of resonant drift of the scroll wave
turbulence is yet to be developed. Yet we suppose that it is $\omegac$
that is to be compared to the forcing frequency to determine
resonance, since it represents the \textit{de facto} state of the
controlled system regardless of the detailed mechanisms that brought
it into that state.
\end{itemize}
The frequency $\omegas$ was measured for a single straight scroll. The
frequencies $\omegat$ and $\omegac(A,\omegaf)$ were both measured by
recording the intervals $T_j$ between the moments $\tr_j$ in which
wavefronts passed through a recording point $(\xr,\yr,\zr)$, that is,
$T_j=\tr_{j+1}-\tr_j$, $u(\xr,\yr,\zr,\tr_j)=\uc$, where $\uc=0$, and
$\df{u}{t}(\xr,\yr,\zr,\tr_j)>0$.  The mean frequency for the entire
simulation was then calculated as the average, $\omegac=2\pi/\langle
T_j\rangle=2\pi N/\sum_{j=1}^NT_j$, for all $N$ intervals recorded,
and similarly for $\omegat$. For $\omegat$ the simulation was run for
$t\in[0,5000]$.  For $\omegac(A,\omegaf)$ the simulation was run for
$t\in[0,5000]$ or until all scrolls were terminated if it happened
sooner.

For our crude discretization steps $\dx=2/3$, $\dt=1/30$ and $L=40$,
when modulating the medium's excitability we have observed
$\omegas=0.90$, $\omegat=1.05$ and $\omegac(0.03,\omegas)=0.74$. The
finer discretization steps $\dx=0.40$, $\dt=0.01$ and $L=60$ produced
$\omegas=1.20$, $\omegat=1.27$ and $\omegac(0.03,\omegas)=1.01$. That
is, the cruder discretiazation slows down the scroll waves overall,
compared to the finer discretization, but the relationship between the
key frequencies remains similar.

\subsection{Forcing}

We investigated the application of the following types of forcing on
scroll wave turbulence;
\begin{enumerate}
  \item Modulation of the medium's excitability, \ie\ variation of
parameter $b(t)$ around its average value $b_0$.
  \begin{enumerate}
    \item Repetitive stimulation
    \begin{enumerate}
      \item Constant frequency
    \end{enumerate}
  \end{enumerate}
\item Extra transmembrane current, $h(t)$
  \begin{enumerate}
    \item Single pulse
    \item Repetitive stimulation
    \begin{enumerate}
      \item Constant frequency
      \item Feedback controlled 
    \end{enumerate}
  \end{enumerate}
\end{enumerate}

\subsubsection{Modulation of the medium's excitability} 

The medium's excitability in the model is defined by parameter
$b$. Following \cite{Alonso-etal-2003,Alonso-etal-2006-Chaos}, we
introduced into the model a spatially-uniform forcing by applying a
periodic modulation of the parameter $b$ in time, while keeping the
extra current term zero,
\begin{eqnarray}
b(t) & = & b_0 + A\cos(\omegaf t), \qquad h(t)=0,  \label{b}
\end{eqnarray}
where $b_0=0.19$, $A$ is the forcing amplitude, and $\omegaf$ is the
forcing frequency. Since $b$ determines the excitation threshold,
varying its value will vary the excitability of the medium.

Starting from the five different turbulent initial conditions, the
simulations were performed at different values of $A$ and $\omegaf$,
and the time taken for elimination of the turbulence was recorded.  If
any scrolls remained after $t=5000$ then the experiment was stopped
and considered to have failed to eliminate the turbulence.

\subsubsection{Extra transmembrane current}

Here, simulations were performed with
\[
b(t) = b_0 = 0.19, \qquad h(t) \ne 0.  
\]

\paragraph*{(a) Single pulse.}

Simulations were started from the saved initial conditions at $t=t_0$
and a single pulse of time duration $\duration=0.3$ was applied,
\[
  h(t) = A \, \Heav(t-t_0) \, \Heav(t_0+\duration-t) ,
\]
where $\Heav()$ is the Heaviside step function.  After the shock,
evolution of the filaments was observed for a further 250 units of
time. If no filaments remained at the end of this period, the shock
was considered to be a success, and otherwise it was deemed to be
unsuccessful.

Shocks of different amplitudes $A$ were tested, and the success
threshold was defined as the amplitude which gives a $50\%$ success
rate over the five initial conditions used.

\paragraph*{(b) Repetitive stimulation.}

The stimulus $h(t)$ was set to be a repetitive series of rectangular
pulses of amplitude $A$,
\[
  h(t) = A \, \sum\limits_{j=1}^{N} \Heav(t-\ts_j) \, \Heav(\ts_j+\duration-t) .
\]
We studied the effect of repetitive stimulation with both constant and
feedback-controlled frequencies.

\begin{itemize}

\item[i.] \textbf{Constant frequency}

  Simulations were started from the saved initial conditions at
  $t=t_0$, and periodic pulses were applied,
  $\ts_j=t_0+j\,2\pi/\omegaf$, $j=0,1,2\dots$.  Experiments were
  repeated for different forcing frequencies $\omegaf$ and amplitudes
  $A$, and the time taken to eliminate the turbulence in each
  experiment was recorded. If any scrolls remained after $t=5000$ then
  the experiment was stopped and considered to have failed to
  eliminate the turbulence.

\item[ii.] \textbf{Feedback controlled}

  Stimulation $h(t)$ was set to be a repetitive series of rectangular
  pulses with timings $\ts_j$ determined by taking feedback from the
  turbulence itself. Feedback is taken from a recording point
  $(\xr,\yr,\zr)$ so that a pulse is applied every time that a wave
  passes through this point, that is, $\ts_j=\tr_j+\tdelay$,
  $u(\xr,\yr,\zr,\tr_j)=\uc$ and $\df{u}{t}(\xr,\yr,\zr,\tr_j)>0$. We
  also varied the time delay $\tdelay$ for
  applying a pulse after a wavefront has passed through the recording
  point. Delays ranging from $\tdelay=0.0$ to $\tdelay=6.0$ were used,
  with increments of $0.1$.

\sglfigure{
  \includegraphics{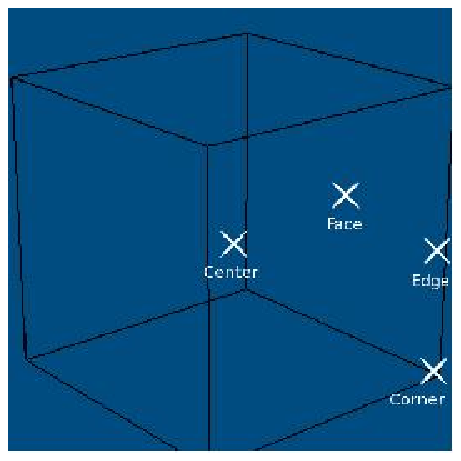}
}{(color online)
  Different locations for the recording point
}{recordingpoint}

Different locations were used for the recording point (see
\fig{recordingpoint}):
\begin{itemize}
\item in the \textbf{corner} of the domain, $(\xr,\yr,\zr) = (0,0,0)$,
\item in the \textbf{center} of the domain, $(\xr,\yr,\zr) = (L/2,L/2,L/2)$,
\item in the center of a \textbf{face} of the domain, $(\xr,\yr,\zr) = (L/2,L/2,0)$,
\item in the center of an \textbf{edge} of the domain, $(\xr,\yr,\zr) = (L/2,0,0)$.
\end{itemize}
Experiments were repeated for different locations for the recording
point, different values of time delay $\tdelay$, and amplitude $A$ and
the time taken for the elimination of the turbulence recorded. If any
scrolls remained after $t=5000$ time steps, then the experiment was
stopped and considered to have failed to eliminate the turbulence.
\end{itemize}

\section{Results}

\subsection{Elimination of the turbulence}

\dblfigure{
  \includegraphics{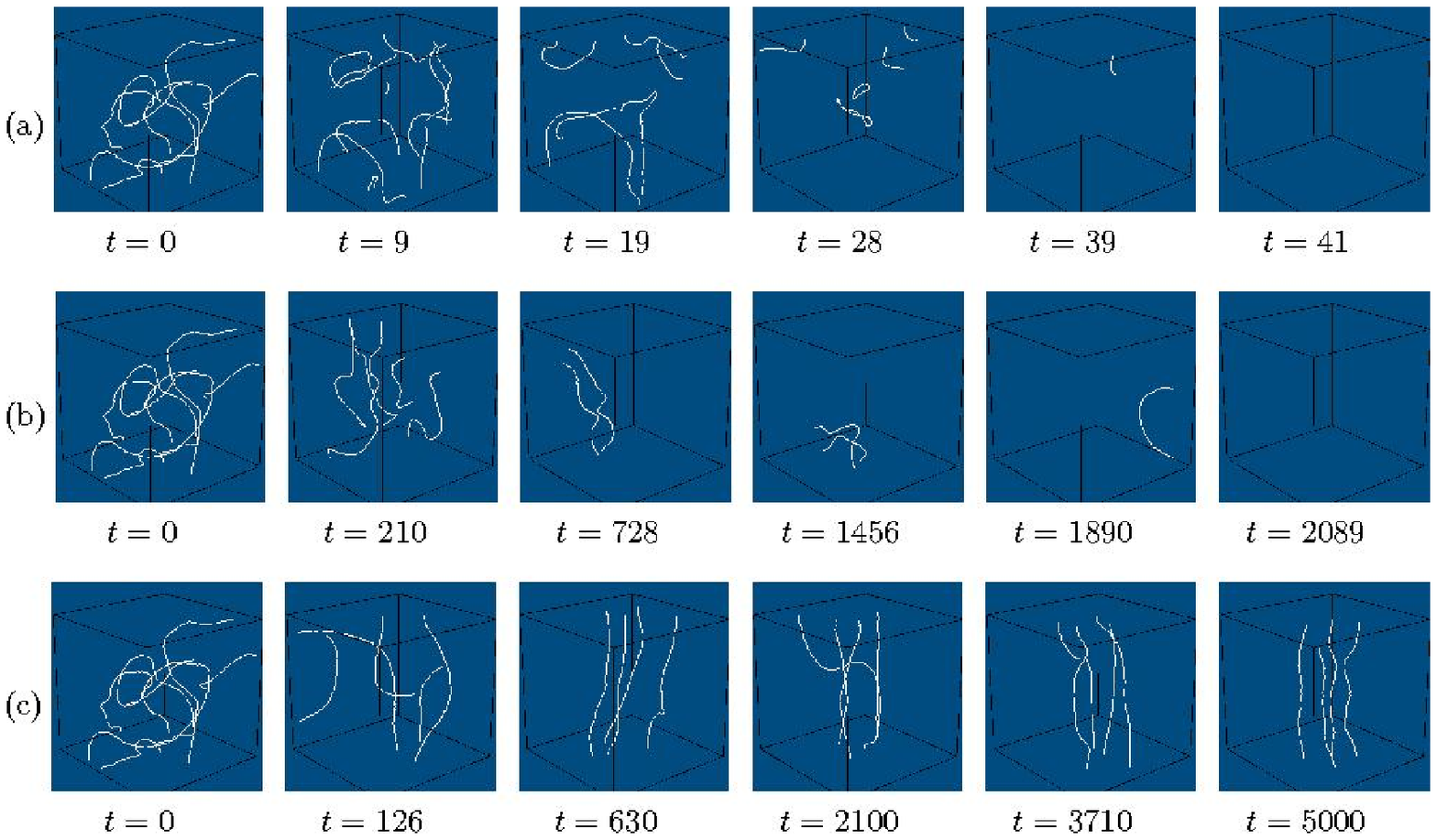}
}{(color online)
  Examples of successful and unsuccessful elimination of turbulence.
  Evolution of the turbulence under modulation of medium excitability with
  (a) $\omegaf=0.8$, (b) $\omegaf=1.22$, (c)
  $\omegaf=1.13$ at fixed amplitude $A=0.03$.
}{ModFigs}

\Fig{ModFigs} illustrates the main result of our study: a resonant
stimulation can eliminate scroll wave turbulence, and does it quicker
and more reliably than a non-resonant stimulation. The figure shows
evolution of the turbulence due to modulation of the medium's
excitability at three different forcing frequencies at a fixed
amplitude. The forcing frequency $\omegaf=0.8$ is within the
``resonant window'' (see below for a formal definition), and the
turbulence is terminated quickly within $t=41$. The forcing frequency
$\omegaf=1.22$ is above-resonant, and although the turbulence is
terminated, it takes much longer, $t=2089$. The forcing frequency
$\omegaf=1.13$ is also above the resonant window, and it leads to
stabilisation of vortices in the center of the medium rather than
their termination.

We note here that the mechanism of ``taming'' of scroll wave
turbulence suggested by Alonso \etal\
\cite{Alonso-etal-2003,Alonso-etal-2006-Chaos} is based on inversion
of the filament tension from negative to positive. The sequence shown
in \fig{ModFigs}(c) illustrates why this is \emph{not} sufficient for
defibrillation: the scroll filaments stabilise with a straight shape,
which is consistent with their effective tension being positive, but
it does \emph{not} lead to their elimination.

In the following subsections, we analyse in more details the empirical
conditions required for successful termination of the turbulence.

\subsection{Windows of resonant frequencies}

For both the modulation of the medium's excitability and extra
transmembrane current forcing, we have varied the frequency $\omegaf$
and amplitude $A$ to assess their effects on termination of the
turbulence. We observed very different effects of the modulation of
excitability and of the extra current forcing on the mean frequency of
the turbulence $\omegac(A,\omegaf)$. In this section we define a
resonant window of frequencies for each amplitude and forcing type.

\subsubsection{Modulation of excitability}\label{ResWindowMOD}

As the amplitude of the modulation increases, the mean frequency of
the turbulence $\omegac(A,\omegaf)$ decreases dramatically (see
\fig{figfreq}(a)). When using the largest forcing amplitude
($A=0.05$), the frequency of the turbulence reduced to over half that
of the frequency of the unperturbed turbulence.

Resonant windows can be identified in \fig{figfreq}(c). We define the
resonant window to be the range of forcing frequencies $\omegaf$, for
which $\omegaf \approx \omegac(A,\omegaf)$.  The upper and lower
bounds for the resonant window can be seen as the first and last
points where $\omegaf - \omegac(A,\omegaf) = 0$. The resonant window
is taken to be this range and a further $0.01$ either side of this
range.

There is a different resonant window for each amplitude. As the
amplitude increases, the size of the resonant window increases and
shifts towards lower forcing frequencies.

The above definition of the resonant window should be used with
caution for the lowest amplitudes. E.g. for $A=0.01$ in
Fig.\ref{figfreq}(c) the window must be between $0.873$ and
$1.045$. However, most of this interval corresponds to a ``false
resonance'', when $\omegaf\approx\omegac$ but that does not lead to
termination. Termination of the turbulence at this forcing amplitude
is observed in the narrow vicinity of $\omegaf=0.873$ only. Applying a
forcing with frequency in the vicinity of $\omegaf=1.045$ maintains
the turbulence.

\subsubsection{Extra transmembrane current}

As the amplitude of the extra current forcing increases, the mean
frequency of the turbulence $\omegac(A,\omegaf)$ decreases (see
Fig.\ref{figfreq}(b)), though not so dramatically as in the case of
modulation of the medium's excitability, compare Fig.\ref{figfreq}(a)
and (b). Even for the largest forcing amplitude $A=0.5$, the reduction
in the frequency of the turbulence due to the extra current forcing is
not more than $20\%$ of the frequency of the unperturbed turbulence.

Resonant windows for extra current forcing can be identified in
Fig.\ref{figfreq}(d). The resonant windows were defined in the same
way as in section \ref{ResWindowMOD}. There is a different resonant
window for each amplitude, although for all forcing amplitudes that we
tested, the resonant windows are in the vicinity of the frequency of a
straight scroll $\omegas=0.9$.

For the lowest amplitude $A=0.1$ in Fig.\ref{figfreq}(d) there is no
obvious resonant window. However, the resonant termination of the
turbulence at this forcing amplitude is observed in the narrow
vicinity of $0.870$. The interval of frequencies above that and up
until $1.09$ corresponds to the ``false resonance''.

\subsection{Termination times}

\subsubsection{Modulation of the medium's excitability}

\dblfigure{
  \includegraphics{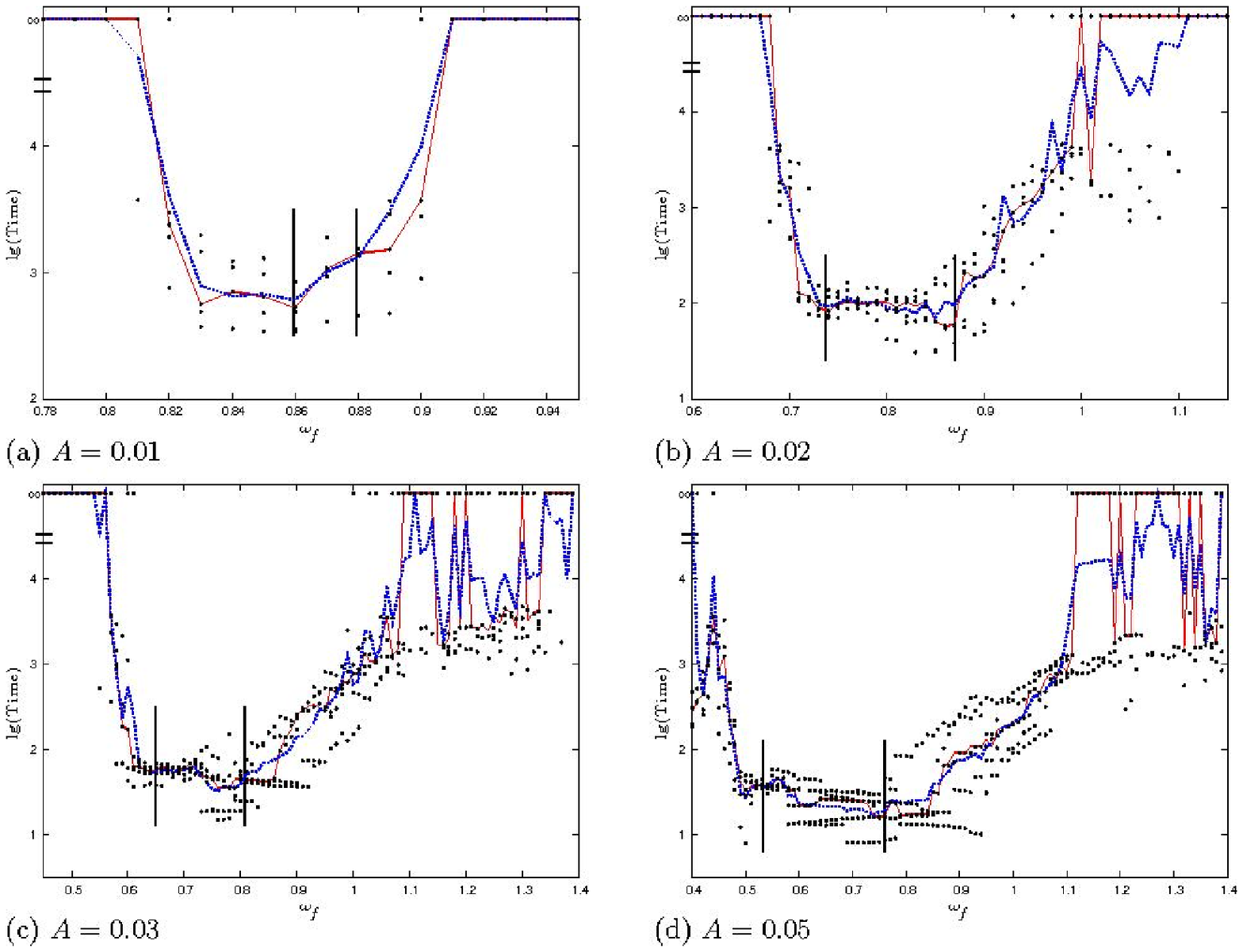}
}{ (color online)
  Modulation of medium's excitability: termination times at
  different amplitudes: (a) $A=0.01$, (b) $A=0.02$, (c) $A=0.03$ and (d)
  $A=0.05$. Black dots: termination times for individual
  simulations. Red solid line: median values of the termination times at
  every fixed frequency. Blue dashed line: same, geometric mean
  values. For averaging and visualization purposes, we assign value
  $t=10^5$ to the failures. Black vertical brackets: the windows of
  resonant frequencies for each amplitude.
}{ModTimes}

\Fig{ModTimes} presents the turbulence termination times due to
modulation of the medium's excitability at four different forcing
amplitudes. The vertical brackets show the windows of resonant
frequencies for each amplitude, as in \fig{figfreq}. From these data,
it can be seen that termination of turbulence is fastest when the
forcing frequency $\omegaf$ is within the resonant window. The
turbulence can be terminated with a frequency outside of the resonant
window.  Although, further away from the resonant window, average
termination time increases and the probability of success decreases.

\dblfigure{
  \includegraphics{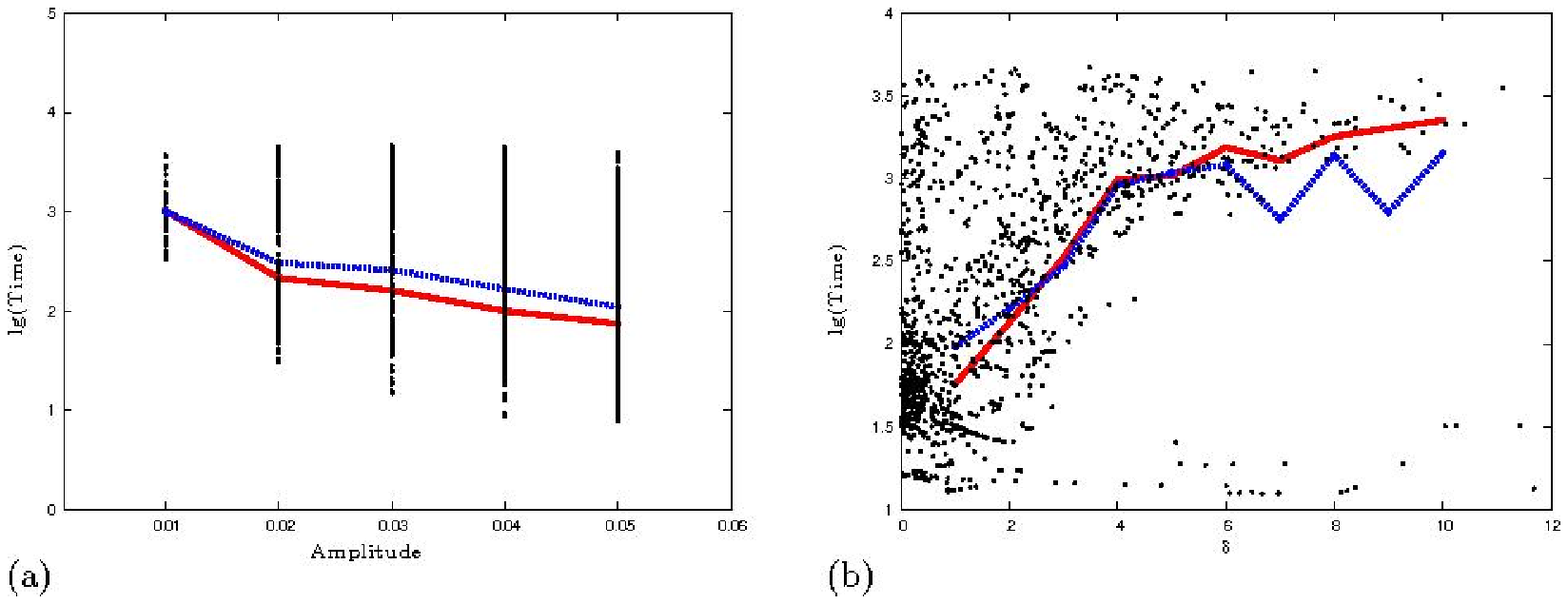}
}{(color online)
  Modulation of medium's excitability: Termination time against
  (a) fixed forcing amplitude $A$, (b) untuning of the resonance
  $\untuning$. The dots represent termination times for individual
  simulations. The solid line goes through the median values of the
  termination times at either (a) or (b), and the dashed line through
  the geometric mean values.
}{F1}

\Fig{F1} shows a plot of termination time against (a) fixed forcing
amplitude $A$, and (b) untuning of the resonance $\untuning$ defined
as:
\[
  \untuning = \frac{|\omegaf - \omegac(A,\omegaf)|}{\sigma}, \label{untuning}
\]
where $\sigma$ is the standard deviation of the turbulence
frequency recorded throughout a simulation, 
\[
  \sigma = 2\pi \left(
    \frac1N \sum_{j=1}^N T_j^{-2}
    -
    \left( \frac1N \sum_{j=1}^N T_j^{-1} \right)^2
  \right)^{1/2} .
\]
The strength of resonance is a measure of how close the forcing
frequency is to the frequency of the scroll wave turbulence.

It can be seen from \fig{F1}(a) that increasing $A$ reduces the
termination time. The reduction is not very pronounced at larger $A$;
one should bear in mind here that the data in this graph are for
\emph{all} frequencies, resonant or not. \Fig{F1}(b) shows that at
smaller $\untuning$, \ie\ a better resonance, the time taken to
eliminate the turbulence reduces. Here the data are for all forcing
amplitudes, large and small. Comparing \fig{F1}(a) and \fig{F1}(b),
and taking into account that reliable forcing amplitudes for the the
turbulence termination seems to be $A \ge 0.02$, termination times are
more sensitive to the quality of resonance than to the forcing
amplitude.

\subsubsection{Extra transmembrane current forcing}

\paragraph*{(a) Single pulse.}

\sglfigure{
  \includegraphics{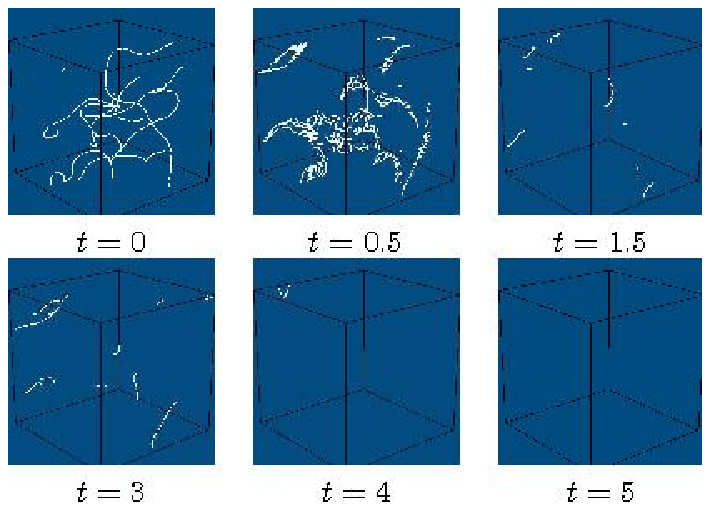}
}{(color online)
  Single pulse: termination of the turbulence at forcing
  amplitude $A=4.3$.
}{SinglePulse}

The single pulse stimulation was tested having in mind the current
clinical practice for cardiac defibrillation is by means of a single
electric pulse of large amplitude. We defined the single pulse success
threshold as the amplitude at which the turbulence is terminated in
more than $50\%$ of the experiments.  In our setup, this success
threshold was found to be $A=4.3$. \Fig{SinglePulse} shows an example
of successful defibrillation with a single pulse shock at amplitude
$A=4.3$.

\paragraph*{(b) Repetitive pulses.}

\dblfigure{
  \includegraphics{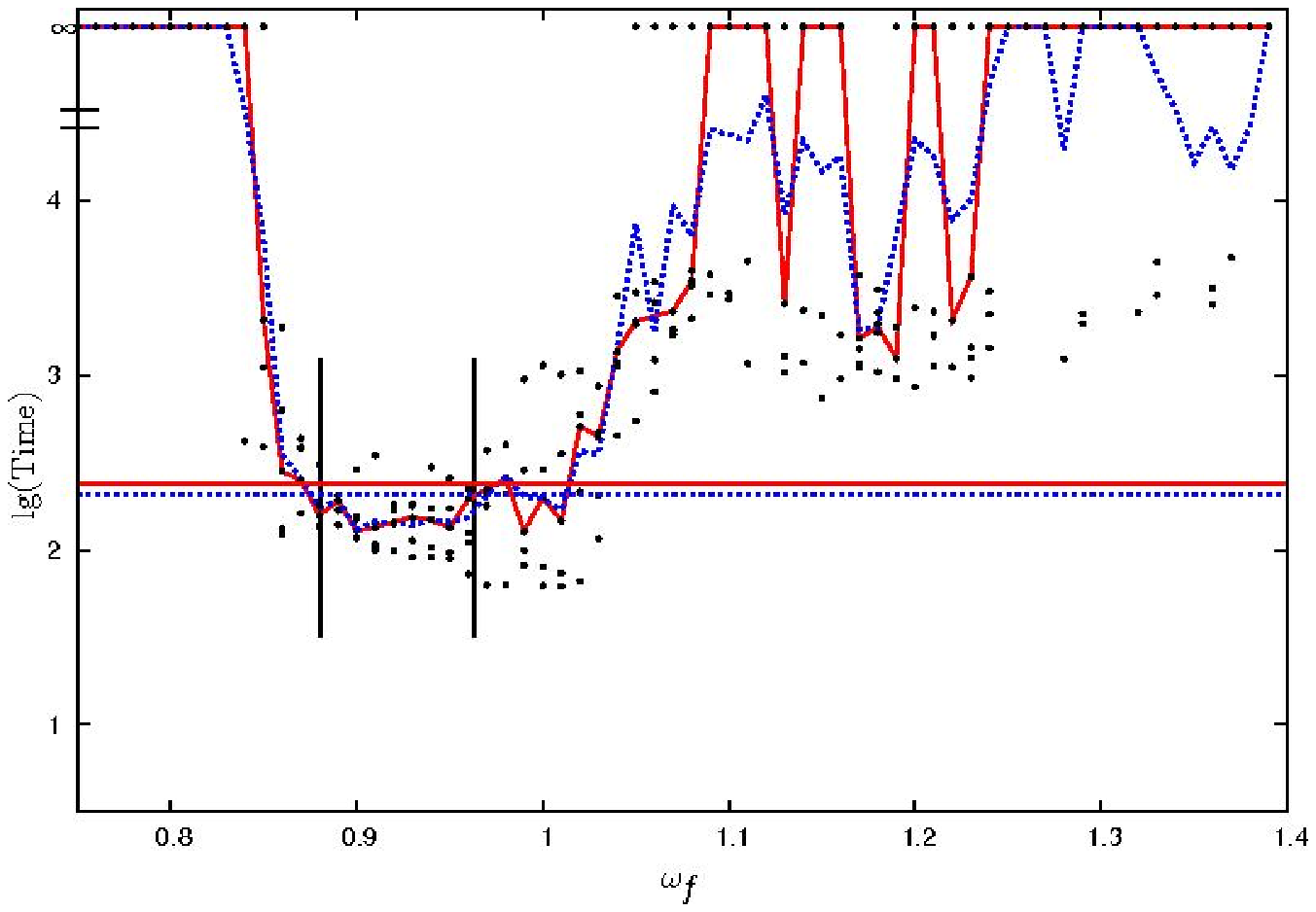}
}{ (color online) 
  Extra current forcing: termination time for the amplitude
  $A=0.3$. The black dots represent termination times for individual
  simulations. The red solid line goes through the median values of the
  termination times at a fixed frequency, and the blue dashed line
  through the geometric mean values. The horizontal straight lines show
  the median (red solid line) and geometric mean (blue dashed line)
  termination times for the feedback experiments. The vertical brackets
  designate the window of resonant frequency.
}{FB03}

\Fig{FB03} presents the turbulence termination times, both for
constant-frequency and feedback-controlled extra currrent forcing. The
solid-line and dashed-line curves show the dependence of the
termination times on the extra current forcing frequency $\omegaf$, at
fixed amplitude $A=0.3$.  The vertical brackets show the window of
resonant frequencies as in \fig{figfreq}(d). The dashed horizontal
straight line shows the geometric mean termination time for the
feedback controlled stimulation, and the solid horizontal line shows
the corresponding median termination time. From these data, it can be
seen that forcing frequencies within the resonant window ensure the
fastest termination of turbulence. The turbulence can be terminated
with a frequency outside of the resonant window but there the
probability of the turbulence termination decreases. The individual
experiments with failed termination are depicted as
$\log_{10}(t)=\infty$, and counted for the sake of averaging as
$\log_{10}(t)=5$. Further away from resonant frequencies the
turbulence termination time rapidly increases.

\dblfigure{
  \includegraphics{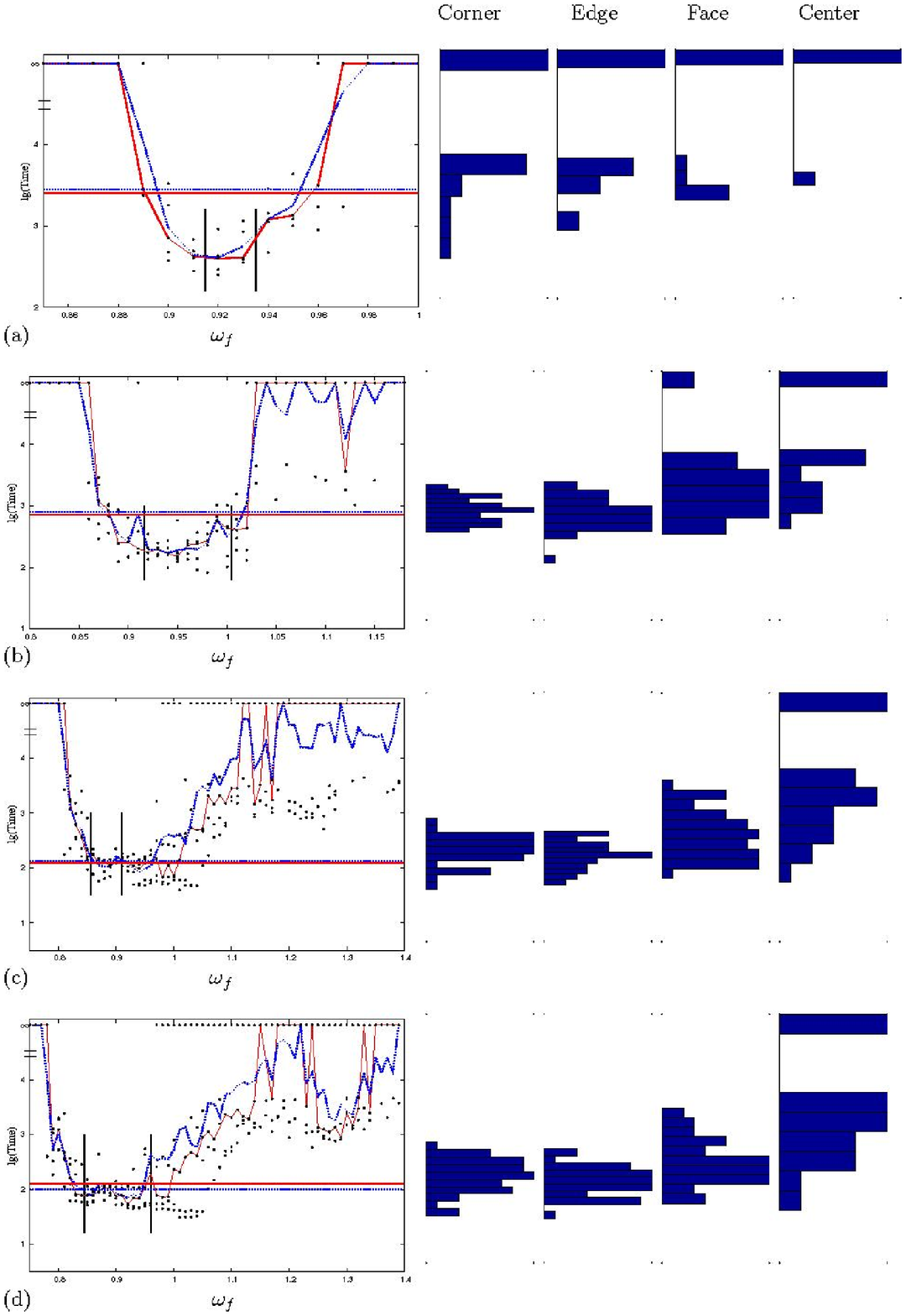}
}{(color online) 
  Extra current forcing: termination times for the amplitudes
  (a) $A=0.1$, (b) $A=0.2$, (c) $A=0.4$, (d) $A=0.5$. The left panels
  are as in Fig.\ref{FB03}. The histograms on the right are distribution
  of termination times for feedback controlled experiments, at different
  locations for the recording point (they are rotated $90^\circ$
  clockwise and flipped in the vertical direction, to bring their
  abscissa axes in line with the ordinata axes of the graphs on the
  left). 
}{FbTimes}

\Fig{FbTimes} illustrates the effects of the stimulation amplitude
$A$. The left panels in it are similar to \fig{FB03} and present the
turbulence termination time dependence on the extra current forcing
frequency $\omegaf$, at four different amplitudes $A=0.1, 0.2, 0.4,
0.5$. The histograms on the right show the distributions of
termination times for feedback controlled experiments, for the four
different locations of the recording point.  All the observations made
for \fig{FB03} are valid for the forcing amplitudes in \fig{FbTimes}.

For amplitudes $A > 0.2$ in feedback controlled experiments, the
average termination time is close to the average termination time
achieved at resonant frequencies. For lower amplitudes, the success
probability using feedback-controlled controlled stimulation falls
down, in the same way as it does for the constant-frequency forcing.

The location of the recording point used for the feedback is also
important for a successful termination. The most successful locations,
in the 3D experiments, appear to be the corner or the edge of the
medium. The center location for the recording point appears to be the
worst for all forcing amplitudes tested.
  
\dblfigure{
  \includegraphics{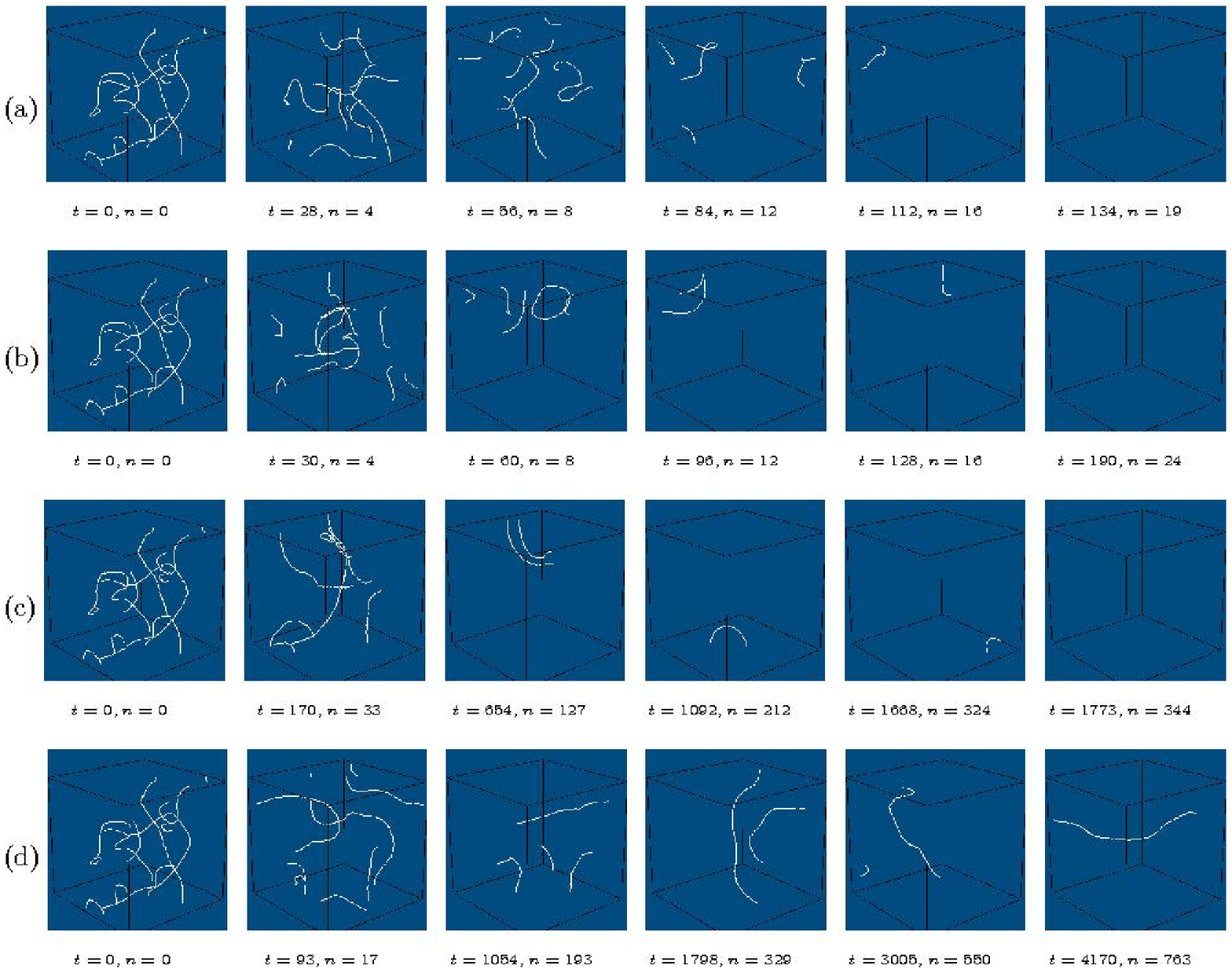}
}{(color online)
  Extra current forcing: evolution of the turbulence using
  (a) $\omegaf=0.9$ (b) feedback (c) $\omegaf=1.22$ (d)
  $\omegaf=1.15$ at fixed amplitude $A=0.3$.
}{ExtImages}

\Fig{ExtImages} shows evolution of the turbulence due to extra current
forcing at different forcing frequencies $\omegaf=0.9, 1.22, 1.15$,
and the feedback controlled, at the fixed amplitude $A=0.3$. The
forcing frequency $\omegaf=0.9$ is within the resonant window, and the
turbulence is terminated quickly by $t=134$ (series a). The feedback
controlled stimulation terminates the turbulence by $t=190$ (series
b). The forcing frequency $\omegaf=1.22$ is above-resonant, although
the turbulence is terminated it takes ten times longer, to $t=1773$
(series c). The forcing frequency $\omegaf=1.15$ is also above the
resonant window. Simulation with that frequency leads to stabilisation
of a vortex in the center of the medium (series d), similar to what
was observed for faster-than-resonant modulation of excitability, see
\fig{ModFigs}(c).

\dblfigure{
  \includegraphics{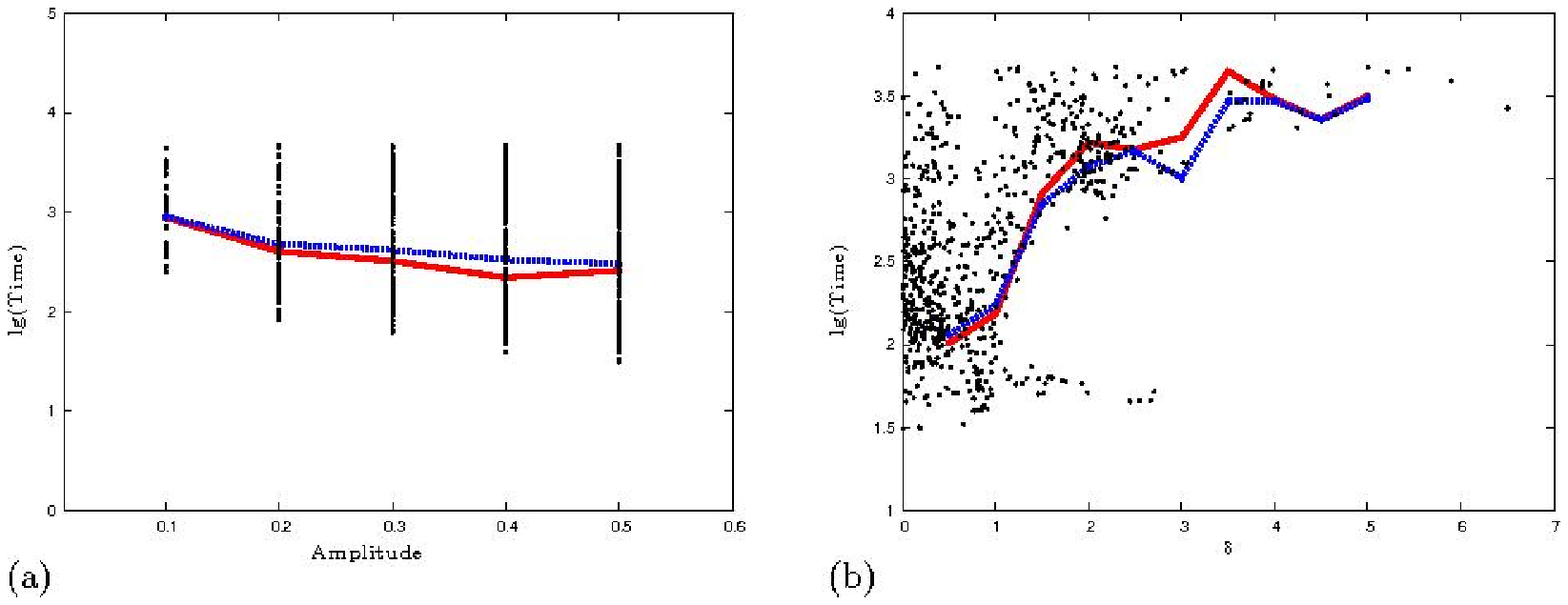}
}{(color online)
  Extra current forcing:
  Termination time against (a) fixed forcing amplitude $A$, (b)
  untuning of the resonance $\untuning$. The dots represent
  termination times for individual simulations. The solid line goes
  through the median values of the termination times at either (a) or
  (b), and the dashed line through the geometric mean values.
}{F2}

\Fig{F2} is similar to \fig{F1} and shows plots of termination time
against (a) fixed forcing amplitude $A$, and (b) untuning of the
resonance $\untuning$, defined in the same way as for the modulation
of excitability forcing. It can be seen from \fig{F2}(a) that
increasing the amplitutde $A$ reduces the termination
time. \Fig{F2}(b) shows that reducing the untuning $\untuning$ also
reduces the termination time. Comparing \fig{F2}(a) and \fig{F2}(b),
and taking into account that reliable forcing amplitudes for the
turbulence termination seems to be $A \ge 0.2$, termination times are
more sensitive to the untuning of resonance $\untuning$ than to the
change of forcing amplitude $A$.

\section{Discussion}

\subsection{Resonant stimulation terminates scroll turbulence}

We have shown that a termination of the scroll wave turbulence can be
achieved by repetitive stimulation. This is despite the fact that
scroll waves continue to grow and multiply while we drive them to
elimination.  We have studied two different methods of forcing, the
modulation of medium's excitability and the extra transmembrane
currents forcing.  Where comparable, the qualitative depenencies for
the two methods were similar.  For a successful termination, the
amplitude of the repetitive forcing should be higher than a certain
threshold. However, this threshold is still much lower than that
required for termination by a single shock. The termination is
achieved with the highest probability, and in the quickest time, when
using a resonant forcing frequency, \ie\ when the frequency of
stimulation, $\omegaf$, is close to the \textit{de facto} frequency of
the forced scroll waves, $\omegac(A,\omegaf)$. Namely, we have shown
that for both types of forcing, the turbulence termination becomes
faster for smaller values of the untuning of the resonance $\untuning$
(see \fig{F1}(b) and \fig{F2}(b)).

\subsection{Resonant windows}

We have found that the resonance between the forcing and the scrolls
is characterised not by a single resonant frequency, but by a resonant
window which depends on the type of forcing and its amplitude. For the
modulation of the medium's excitability, larger amplitudes
corresponded to wider resonant windows, shifted towards lower forcing
frequencies. For the extra transmembrane current forcing, the resonant
windows did vary in size and location, but for the amplitudes that we
tested they remained close to the frequency of a single straight
vortex, $\omegas\approx0.9$.

Our results also show that for both modulation of excitability and
extra current forcing, the fastest termination was achieved when the
forcing was applied with a frequency chosen within the resonant
window. Termination using a forcing frequency outside of the resonant
window occurs with a much lower probability and a longer termination
time.

Existence of resonant windows rather than unique resonant frequencies
may be a purely statistical phenomenon due to fluctuations of
frequencies of the scroll turbulence, or may be an indication that the
forced turbulence adjusts its frequency in response to the forcing,
\ie\ a ``frequency locking'' in average (perfect locking is not
feasible between a periodic forcing and chaotic turbulence). Nothing
like this has been reported for resonant drift in 2D, thus we may be
dealing with a specifically three-dimensional phenomenon. Indeed, 3D
scrolls have additional degrees of freedom compared to 2D spirals,
\eg\ twisted scrolls can rotate faster than straight scrolls
\cite{Panfilov-etal-1984,Mikhailov-etal-1985}.  This might offer an
explanation of cases of false resonance, when formally defined
resonant windows are abnormally extended towards higher frequencies
but are not associated with fast and reliable termination, see \eg\
the case of $A=0.01$ in \fig{figfreq}(a,c) and
\fig{ModTimes}(a). Indeed, in case of strong twist, different parts of
the same filament have different phases and are forced in different
directions, which does not result in an overall directed movement and
does not bring about termination.

\subsection{Resonant vs non-resonant stimulation}\label{subsec:res-v-nonres}

It has been previously shown \cite{Alonso-etal-2003} in Barkley's
model with the same model parameters as we used here, that scroll wave
turbulence can be controlled by a weak \emph{non-resonant} modulation
of the medium's excitability. A theory was presented in
\cite{Alonso-etal-2006-Chaos} explaining that this control of
turbulence was due to an inversion of the filament tension from
negative to positive, which can happen if the frequency of forcing is
higher than the frequency of the scrolls. It was argued that such
stimulation causes the filaments to collapse and could therefore be
used for termination of the scroll wave turbulence.  We have seen,
however, that positive tension may lead to stabilization rather than
termination of scrolls, see \fig{ModFigs}(c) and \ref{ExtImages}(d).

In \cite{Alonso-etal-2003} an above-resonant frequency forcing was
used to control the turbulence. More specifically, their forcing
frequency $\omegaf$ was almost equal to ($1\%$ higher than) the
frequency of a straight scroll, $\omegas$.  The frequency of a forced
turbulence is significantly lower than the frequency of a straight
scroll, \eg\ $\omegac(A,\omegas)<\omegas$, thus forcing with frequency
$\omegas$ is above-resonant. This is true both for the finer
discretization steps used in \cite{Alonso-etal-2003} and cruder
discrtetization steps used in a majority of our simulations.

To make a specific comparison, let us consider excitability modulation
with amplitude $A=0.03$ (same as in \cite{Alonso-etal-2003}) and
frequency $\omegaf=0.91$ which is about $1\%$ higher than
$\omegas$. As can be seen from \fig{ModTimes}(c), such forcing gives a
mean turbulence termination time of $t=251$ (compare with termination
time $t=1510$ in the example shown in \cite{Alonso-etal-2003}). Within
the resonant window for this amplitude as defined in this article, the
mean termination time is between $t=71$ and $t=32$, \ie\ 3 to 7 times
faster than using above-resonant forcing frequency as in
\cite{Alonso-etal-2003}.

So, a direct like-for-like comparison shows that although the
above-resonant frequency stimulation suggested in
\cite{Alonso-etal-2003} works in principle, the resonant stimulation
works more reliably and much faster.

It has been reported by Wu \etal~\cite{Wu-etal-2006}
  that using a travelling-wave modulation of the mediums excitability
  can control scroll wave turbulence faster than the modulation used
  in \cite{Alonso-etal-2003}. Fig.6 in
  \cite{Wu-etal-2006} clearly shows that the optimum forcing frequency
  is below the rotation frequency of an unforced vortex $\omegas$.
  Alas, Wu \etal\ did not control the \textit{de facto} frequencies of
  the scroll turbulence so it is not possible to interpret their
  results unambiguously. However, as we have shown here that the
  resonant frequencies are below $\omegas$, it is quite possible
  that the real reason for the advantage achieved in \cite{Wu-etal-2006}
  compared to \cite{Alonso-etal-2003}
  is not (only) in using travelling waves, but
  simply in using a frequency within a resonant window. A definitive
  answer to this question requires further investigation.

\subsection{Feedback control works in 3D}\label{subsec:fb3d}

As the resonant window may not be known \textit{a priori}, we tested a
simple algorithm for the feedback control with the extra current
forcing. It has been previously shown in 2D that applying a repetitive
feedback controlled forcing causes a spiral wave to drift to a
boundary along a predictable trajectory
\cite{Biktashev-Holden-1994,Biktashev-Holden-1995}, and possibly
terminating it faster than with the constant frequency forcing, as
feedback forcing can overcome the ``resonant repulsion'' of the
drifting spirals from boundaries and inhomogeneities
\cite{Biktashev-Holden-1993}.  Simulations of spiral waves in
  the two-dimensional version of our model easily demonstrate resonant
  repulsion (see \fig{2dsquiggle}(a)), so it must play some role in 3D behaviour as well, even
  though it may be not straightforward in scroll turbulence.%
Our 3D experiments show that the termination
with a feedback-controlled forcing is nevertheless achievable even at relatively
weak amplitudes.

Our results also show that the location of the recording point is
important for a successful feedback-controlled termination, 
which is in good agreement with earlier observations of feedback-driven
resonant drift in two dimensions~\cite{%
  Nikolaev-etal-1998,%
  Panfilov-etal-2000%
}. The most
successful locations, in the 3D experiments, appear to be the corner
or the edge of the medium. The center location for the recording point
appears to be the worst for all forcing amplitudes tested. It has been
shown in 2D experiments \cite{Zykov-Engel-2007} that a line of
recording points is a robust approach. Therefore, if the same holds for
3D, a line of recording points down one edge of the medium may be the
optimal choice.

\dblfigure{
  \includegraphics{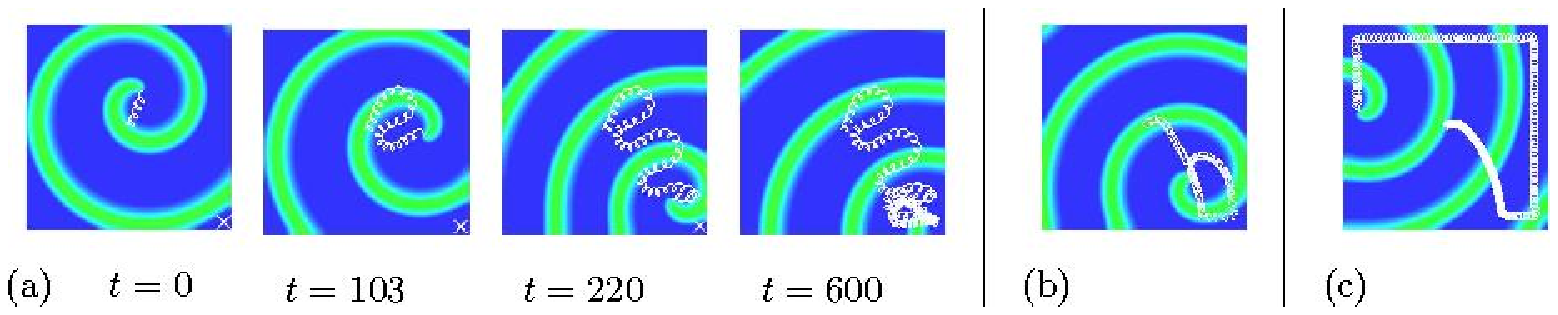}
}{ (color online)
  Two-dimensional phenomena.
  (a) `Resonant reflection': 
  Extra current forcing with constant frequency $\omegaf=1.1855$ and amplitude $A=0.01$.
  (b-d) `Snake instability': Extra current forcing
  with feedback controlled frequency applied with amplitudes
  $A=0.01$ (b), $A=0.004$ (c)  and $A=0.001$ (d).  Model parameters $a=1.1$,
  $b=0.19$, $\epsilon = 0.02$, $L=60$,
  $\dx=0.4$ and $\dt=0.01$, with 9-point approximation of the Laplacian.
  The recording point in (b--d) was located in the
  bottom left corner.
}{2dsquiggle}

\subsection{Feedback controlled vs constant frequency stimulation}\label{subsec:fb-v-cf}

Typically, the average termination time using feedback is close to
that achieved at resonant constant frequencies.  At the lowest
stimulation amplitudes, both feedback-controlled and
constant-frequency termination times increase, and the probability of
success decreases.

For a range of frequencies within or near the resonant window, the
termination was achieved, on average, quicker than with feedback. To
elucidate a possible reason for this difference, we have considered a
two-dimensional version of the Barkley model, with the same model
parameters and applied the same feedback controlled forcing as in our
3D simulations. We have found that applying a forcing with feedback in
this case causes the spiral wave to drift along a rather complicated
``snaky'' trajectory, shown in \fig{2dsquiggle}(b). This behaviour is
similar to that described recently by 
Zykov \etal~\cite{Zykov-etal-2002,Schlesner-etal-2008},
and is caused by an instability, related to the
delay between a change of the position or phase of the spiral and its
detection by the feedback electrode, due to the distance between them
and a finite speed of the waves. Indeed, in \fig{2dsquiggle} the
spiral core is more than one wavelength away from the recording point. It
can be seen from \fig{2dsquiggle}(b) at $t=525$ that this resonantly
drifting spiral does not terminate at the boundary. Instead, it
embarks on a continuous loop near the boundary of the medium.

According to \cite{Zykov-etal-2002,Schlesner-etal-2008},
this `snake instability' should disappear at lower amplitudes. Indeed,
this is what happened in our simulations, see \fig{2dsquiggle}(c,d).
The instability was less pronounced when the amplitude is
  decreased from $A=0.01$ in (b) to $A=0.004$ in (c), and was
  completely gone for $A=0.001$ in (d). However, although the drift
  trajectory towards the boundary was shorter, the drift velocity was
  smaller proportionally to $A$, so the time to reaching the boundary
  did not decrease. Besides, the decrease in $A$ made annihilation at
  the boundary even less likely: in (c) the spiral stuck near the
  boundary similar to (b), whereas at a further reduction of $A$ in (d) it
  embarked on a drift along the boundary which was faster than its
resonant drift in the center of the domain. So, reducing the amplitude
is not necessarily a satisfactory solution to the problem of the
`snake instability' interference with terminating the
vortex. Alternative solution could be to reduce the delay in the
feedback, say by using a global (ECG) rather than local (point
electrogram) signals. However, in known computer models and
experimental observations, scroll wave are apparently rather large in
the scale of the heart chambers and are typically no more than one
wavelength away from a boundary. Thus the `snake instability' may not
be a problem in a real heart.

\subsection*{Acknowledgement}

We are grateful to S.~Alonso for helpful advice on details of
simulations described in
\cite{Alonso-etal-2003,Alonso-etal-2006-Chaos}.  This study was
supported in part by EPSRC grant EP/D500338/1.


\end{document}